\documentclass[reprint,aps,prb,twocolumn,superscriptaddress,showpacs]{revtex4-1}

\usepackage{graphicx}
\usepackage{dcolumn}
\usepackage{bm}
\usepackage{bbold}
\usepackage{natbib}
\usepackage{amsmath}
\usepackage{color}

\begin{document}

\preprint{APS/123-QED}

\title{Aharonov-Bohm and Aharonov-Casher effects in double quantum dot Josephson junction}

\author{Damian Tomaszewski}
\affiliation{Institute of Molecular Physics, Polish Academy of Science, Smoluchowskiego 17, 60-179 Poznan, Poland}
\author{Piotr Busz}
\affiliation{Institute of Molecular Physics, Polish Academy of Science, Smoluchowskiego 17, 60-179 Poznan, Poland}
\author{Rosa L\'opez}
\affiliation{Institut de F\'{\i}sica Interdisciplinar i de Sistemes Complexos IFISC (CSIC-UIB), E-07122 Palma de Mallorca, Spain}
\author{Rok \v{Z}itko}
\affiliation{Jo\v{z}ef Stefan Institute, Jamova 39, SI-1000 Ljubljana, Slovenia}
\affiliation{Faculty  of Mathematics and Physics, University of Ljubljana, Jadranska 19, SI-1000 Ljubljana, Slovenia}
\author{Minchul Lee}
\affiliation{Department of Applied Physics and Institute of Natural Science, College of Applied Science, Kyung Hee University, Yongin 17104, Korea}
\author{Jan Martinek}
\affiliation{Institute of Molecular Physics, Polish Academy of Science, Smoluchowskiego 17, 60-179 Poznan, Poland}

\date{\today}

\begin{abstract}
We analyze a Josephson junction between two superconductors
interconnected through a normal-state nanostructure made of two
parallel nanowires with embedded quantum dots.
We study the influence of interference effects
due to the Aharonov-Bohm (AB) and Aharonov-Casher (AC) phases for local and nonlocal (split) Cooper pairs.
In the AB effect the phase of electron is affected by magnetic flux, while in the AC effect the phase of the electron in solid state can be modified due to the Rashba spin-orbit coupling.
In the low-transmission regime the AB and AC effects can be related to only local or nonlocal Cooper pair transport, respectively.
We demonstrate that by the addition of the quantum dots
the Cooper pair splitting can be made perfectly efficient,
and that the AC phase is different for non-spin-flip and spin-flip transport processes.
\end{abstract}

\maketitle

\section{Introduction}

Spatially separated spin-entangled electrons in solid state are a
crucial element of quantum communication and
computing~\cite{NielsenQI2000}. One proposal for
creating such entangled states is based on Cooper pair splitting~\cite{RecherPRB01,EldridgePRB10,HofstetterNature2009,HerrmannPRL10,SchindelePRL12,DasNC2012,TanPRL2015}. Nonlocal split Cooper pairs can also be observed in the Josephson junction~\cite{DeaconNC15}, as pointed out by Wang and Hu~\cite{WangPRL11} in regard to the Aharonov-Bohm (AB) effect.
In this paper we study the interference properties of this new Josephson current.

In addition to the well-known Aharonov-Bohm (AB) effect~\cite{AharonovPR59,SharvinJETP81,WebbPRL85,vanOudenaardenNature98},
where the phase of a charged particle is affected by magnetic vector
potential, we consider the dual phenomenon, namely the Aharonov-Casher (AC) effect~\cite{AharonovPRL84,CimminoPRL89,Mathur.68.2964,RichterPhysics2012},
in which electric field acts on the phase of magnetic moment. The AC effect for electrons was observed in mesoscopic rings~\cite{KonigPRL06,BergstenPRL06,NagasawaPRL2012},
or in the Datta-Das transistor~\cite{DattaAPL90,KooScience2009}, where oscillations of conductance as a function of electric field occur
due to the Rashba spin-orbit interaction phase~$\phi_R$, which can be controlled by an external gate voltage.

The Rashba spin-orbit interaction~\cite{RashbaSPSS60,BychkovJPC84,ManchonNatMat2015} can be described by the Hamiltonian:
\begin{equation}
H_{\rm{R}} = \frac{\eta}{\hbar}\left(\vec{p}\times\vec{\sigma}\right)_{\rm{y}}\;,
\end{equation}
where $\eta$ is the Rashba parameter and the $\rm{y}$ axis is perpendicular to the 2DEG plane. Restriction of the movement of electrons to the $\rm{x}$ direction ($k_{\rm{z}}=0$), leads to different wave vectors, $k_\uparrow \neq k_\downarrow$, of electrons with the same energy and spin polarizations $\pm \rm{z}$, due to the spin-orbit interaction. This entails different phases of spin-up and spin-down ($\sigma=~\uparrow,\downarrow$) electrons, $\phi_{\sigma}=k_\sigma L = \phi_0 + \sigma \phi_{\rm{R}}$, where $L$ is the length of the transport channel. It can be shown that the phase of a moving electron depends on its spin $\left| \sigma \right \rangle_{\rm{z}} \rightarrow \exp \left(\sigma\phi_{\rm{R}}\right) \left| \sigma \right \rangle_{\rm{z}}$, where we have omitted the common phase factor $\phi_0$.

In the s-wave superconductors, the Cooper pairs are in the singlet
state, with no net magnetic moment (spin $S = 0$), consequently there
should be no AC effect for a Cooper pair. This is due to the fact that the two spin components ($\sigma=\pm1$ for spin $\uparrow, \downarrow$) of a Cooper pair in a quasi-1D quantum wire
have opposite Rashba phases $\sigma
\phi_R$, so that they cancel each other and suppress the AC effect. This obstacle hindering the manipulation of Josephson current by the spin-orbit interaction can be avoided by breaking of the time-reversal symmetry,
e.g. by a magnetic-field-induced Zeeman splitting or by magnetic exchange interactions \cite{BezuglyiPRB02,KriveLTP04,KrivePRB05,DellAnnaPRB07,BuzdinPRL08,ReynosoPRL08,ZazunovPRL09,BrunettiPRB13,YokoyamaPRB14,SamokhvalovSR14,JacobsenPRB15,MironovPRL15,CampagnanoJPCM15}.

The desired spin control without breaking
of the time-reversal symmetry can be achieved, however, for split
nonlocal Cooper pairs~\cite{TomaszewskiPRB2018}. Since each
electron in the singlet state has a magnetic moment related to its
spin ($S = 1/2$), it is possible to induce the AC effect for each
electron of the pair separately, when a Cooper pair is split and nonlocally preserves its entangled singlet state.

One of the methods to obtain separated Cooper pairs is the use of
double quantum dot both in the
Y-junctions~\cite{RecherPRB01,EldridgePRB10,HofstetterNature2009,HerrmannPRL10,SchindelePRL12,DasNC2012,TanPRL2015}
and the Josephson
junction~\cite{DeaconNC15,ChoiPRB00,PanPRB2006,WangPRL11,JacquetPRB2015,ProbstPRB2016}.
In this double quantum dot~(DQD) system Cooper pairs splitting is
made possible by having the electrons repel each other by strong Coulomb interaction, $U\rightarrow\infty$.

In this paper, we want to analyze whether the presence of quantum dots
in the Joesphson junction leads to a change in the interference
properties of the system. Our calculations show that the AB and AC
effects are still linked to local and nonlocal (split) Cooper
pair transport, respectively. However, the nonlocal component of the
Josephson current has a Rashba phase dependence that depends on
the ground state of the DQD. In this system, in the singlet ground
state, one can observe non-spin-flip and spin-flip transport processes
related to different AC phases, which results in beating in the AC
effect.
We further investigate how the presence of quantum dots influences the Cooper pair splitting efficiency.
We show that also out of the resonant position of the quantum dots energy levels, in cotunneling regime, the splitting efficiency is relatively high.
It can also be shown that due to the Rashba spin-orbit
interaction, there is a possibility to create and tune admixture of triplet
correlations $T_0$ on the quantum dots.
In the last section of this paper, we propose method of
experimental confirmation of our predictions based on the critical Josephson current oscillations as a function of Rashba phase.

\section{Model}

\begin{figure}[t!]
\centering
\includegraphics{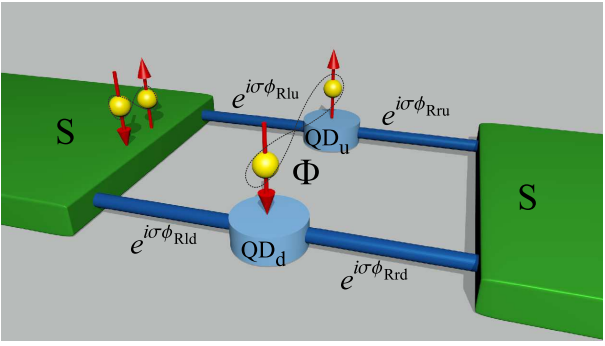}
\caption{Superconducting leads connected by two nanowires with quantum dots (u - up, d - down). Electrons flowing in the system acquire phase $\phi_{\rm{AB}}$ related to the external magnetic field flux $\Phi$ and spin-dependent phase $\sigma \phi_{\rm{R \mu \delta}}$ related to the Rashba spin-orbit interaction. Here $\mu=\left\{\rm{l,r}\right\}$ denotes the left or right section of the nanowire, and $\delta=\left\{\rm{u,d}\right\}$ the up or down nanowire.}
\label{fig:scheme}
\end{figure}

\begin{figure}[t!]
\centering
\includegraphics{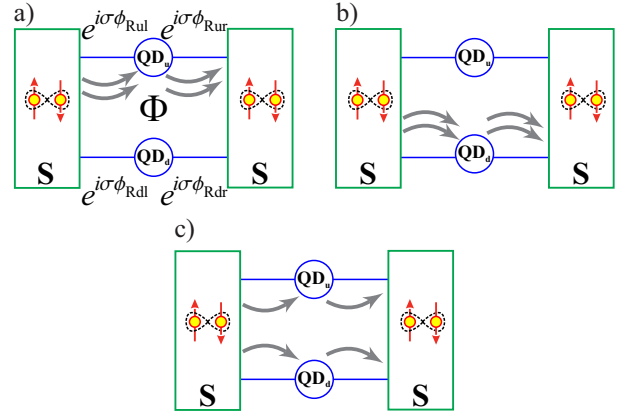}
\caption{Scheme of the 4-th order Cooper pair tunneling processes in the Josephson junction with double quantum dot: a-b) local processes, c) nonlocal process.}
\label{fig:localnonlocal}
\end{figure}

We consider two superconducting leads linked by two parallel nanowires
with en embedded quantum dot in each wire. The distance between
nanowires is smaller than the size $\xi$ of the Cooper pair. In this
system, the flowing electrons acquire Aharonov-Bohm phase
($\phi_{\rm{AB}}$) related to the external magnetic field flux $\Phi$
and spin-dependent Rashba phase ($\sigma \phi_{\rm{R \mu \delta}}$)
related to the Rashba spin-orbit interaction -
Figure~\ref{fig:scheme}. Due to the presence of quantum dots, flow of
the Cooper pairs through the system is dominated by fourth order
processes, schematically shown in Figure~\ref{fig:localnonlocal}. In
processes a) and b) Cooper pair flows through $QD_u$ or $QD_d$ (up or
down quantum dot), respectively, which gives the local contribution to
Josephson current, and in processes c) transport is nonlocal, i.e., the Cooper pair is split between
$QD_u$ and $QD_d$. In the following, we will examine how the presence
of quantum dots affects the interference effects (AB and AC), and we will propose an experimental method that can confirm our assumptions.

The system with a double quantum dot (DQD) inserted between two superconductors
can be described by the Hamiltonian $H = {H_{\rm{l}}} +{H_{\rm{r}}}+ {H_{\rm{DQD}}} + {H_{\rm{Tu}}} + {H_{\rm{Td}}}$,
where the BCS Hamiltonian~$H_\mu$ of superconducting lead~$\mu=\{\rm{l,r}\}$ is given by~\cite{CohenPRL62}:
\begin{equation}
{H_\mu } = \sum\limits_{k\sigma } {{\xi _k}c_{\mu \sigma k}^\dag {c_{\mu \sigma k}} + \sum\limits_k {\left( {{\Delta _\mu }c_{\mu k \uparrow }^\dag c_{\mu  - k \downarrow }^\dag  + H.c.} \right)} }\;.
\label{eqn:HBCS}
\end{equation}
The DQD with single energy level in each quantum dot is described by the Hamiltonian:
\begin{equation}
{H_{\rm{DQD}}} = \sum\limits_{\delta=\rm{u,d}}  {\left( {{\varepsilon _{\delta}}{n_\delta } + {U}{n_{\delta  \uparrow }}{n_{\delta  \downarrow }}} \right)} \;,
\label{eqn:HDQD}
\end{equation}
where $n_{\delta \sigma}$ is the number operator of particles in the $\delta$ quantum dot (u - up, d - down)
and $U$ is the energy of Coulomb interaction.
In our system we set $U\rightarrow\infty$, which implies the impossibility of filling a single quantum dot
with more than one electron.
In the presence of magnetic flux and the Rashba spin-orbit interaction
the tunneling Hamiltonians~$H_{\rm{Tu/d}}$ are given by~\cite{LopezPRB07,CrisanPRB09,WangPRL11}:
\begin{equation}
{H_{\rm{Tu/d}}} = \sum\limits_{k\sigma\mu } {{e^{i\left({\pm\phi _{\rm{AB}}}/4+\sigma \phi _{\rm{R}\mu \rm{u/d} }+\varphi/4\right)}}t_{\mu}d_{\rm{u/d}\sigma }^\dag {c_{\mu k\sigma }}} + H.c.\;,
\label{eqn:HTu}
\end{equation}
where $\varphi$ is the superconducting phase difference, $\sigma\phi_{\rm{R\mu u/d}}$ are the spin-dependent Rashba phases,
$\phi_{\rm{AB}}=\pi \Phi/\Phi_0$ is the AB phase, with
$\Phi_0=h/2e$, $c^\dag_{\mu k\sigma}$ and $d^\dag_{\rm{u/d}\sigma}$ are electron creation operators
on lead~$\mu$ and the $\rm{u/d}$ quantum dot, respectively.
In our model we assume that the Rashba phases
on the left and right side of the two quantum dots ($\phi_{\rm{Rlu/d}}\neq\phi_{\rm{Rru/d}}$)
can be controlled independently. Different Rashba phases can result from different nanowires lengths on the left and right side of QD, or can be controlled by independent gate electrodes.

We consider the system at low temperature in the absence of excited quasiparticles, $\Delta \gg k_BT$, and in the regime $\Delta \gg k_BT_K$, therefore we can neglect the Kondo effect~\cite{ChoiPRB2004}.

The Josephson current $I_J$ can be calculated from the equation:
\begin{equation}
I_J  = \frac{{2e}}{\hbar }\frac{\partial }{{\partial \varphi }}E_{\rm{gs}}\left(\varphi\right)\;.
\end{equation}
Applying the time-independent non-degenerate perturbation theory one can show that first-order and third-order ground state energy shifts are equal $E^{\left(1\right)}_{\rm{gs}}=E^{\left(3\right)}_{\rm{gs}}=0$, the second-order energy shift is independent of superconducting phase difference and will have no contribution to Josephson current, $E^{\left(2\right)}_{\rm{gs}}\left(\varphi\right)=const$. Therefore, the lowest-order energy shift is the fourth-order, which has a form:
\begin{align}
E_{\rm{gs}}^{\left( 4 \right)} =& \left\langle {\rm{gs}} \right|\mathcal{V} \frac{{1 - {P_{\rm{gs}}}}}{{{E_{\rm{gs}}^{\left( 0 \right)}} - {H_0}}}\mathcal{V} \frac{{1 - {P_{\rm{gs}}}}}{{{E_{\rm{gs}}^{\left( 0 \right)}} - {H_0}}}\mathcal{V} \frac{{1 - {P_{\rm{gs}}}}}{{{E_{\rm{gs}}^{\left( 0 \right)}} - {H_0}}}\mathcal{V} \left| {\rm{gs}} \right\rangle  \nonumber\\
-& \left\langle {\rm{gs}} \right|\mathcal{V} \frac{{1 - {P_{\rm{gs}}}}}{{{E_{\rm{gs}}^{\left( 0 \right)}} - {H_0}}}\mathcal{V} \left| {\rm{gs}} \right\rangle \left\langle {\rm{gs}} \right|\mathcal{V} {\left( {\frac{{1 - {P_{\rm{gs}}}}}{{{E_{\rm{gs}}^{\left( 0 \right)}} - {H_0}}}} \right)^2}\mathcal{V} \left| {\rm{gs}} \right\rangle \ ,
\label{eqn:E4}
\end{align}
where the complementary projection operator is defined as $1-P_{\rm{gs}}=1-\left | gs \right \rangle \left \langle gs \right | = \sum_{m \ne {\rm{gs}}}\left | m^{\left(0\right)} \right \rangle \left \langle m^{\left(0\right)} \right |$, and $\mathcal{V}\equiv H_{\rm{Tu}}+H_{\rm{Td}}$ is a perturbative part of the Hamiltonian, which is the only one dependent on the superconducting phase difference.
The last term in Eq.~(\ref{eqn:E4}) can be neglected for calculation of the Josephson current, since it does not depend on the phase difference $\varphi$. As a result:
\begin{align}
&I_J = \frac{{2e}}{\hbar }\frac{\partial }{{\partial \varphi }}E_{\rm{gs}}^{\left( 4 \right)} = \nonumber\\
&=\frac{{2e}}{\hbar }\frac{\partial }{{\partial \varphi }}\left\langle {\rm{gs}} \right|\mathcal{V} \frac{{1 - {P_{\rm{gs}}}}}{{{E_{\rm{gs}}^{\left( 0 \right)}} - {H_0}}}\mathcal{V} \frac{{1 - {P_{\rm{gs}}}}}{{{E_{\rm{gs}}^{\left( 0 \right)}} - {H_0}}}\mathcal{V} \frac{{1 - {P_{\rm{gs}}}}}{{{E_{\rm{gs}}^{\left( 0 \right)}} - {H_0}}}\mathcal{V} \left| {\rm{gs}} \right\rangle \ ,
\label{eqn:IJE4}
\end{align}
where the ground state $\left| {\rm{gs}} \right\rangle  = \left| \rm{BCS} \right\rangle _{\rm{l}} \otimes \left| {\rm{gs}} \right\rangle _{\rm{DQD}} \otimes \left| \rm{BCS} \right\rangle _{\rm{r}}$.

\section{Josephson current}

We consider three DQD ground states: the state~$\left|0\right\rangle$ with empty quantum dots, the state~$\left|01\right\rangle$ where only one quantum dot is occupied by a single electron, and the state where each dot is occupied by a single electron together creating the singlet state~$\left|S\right\rangle$. These ground states can be obtained by tuning the DQD energy, $\varepsilon_{\delta}>0$ corresponding to empty quantum dot, and $\varepsilon_{\delta}<0$ corresponding to the single electron occupation. In the case where the quantum dots are singly occupied we consider only singlet state since it is the ground state of DQD in the strong Coulomb interaction limit~\cite{ProbstPRB2016}. It can also be shown that the nonlocal contribution to the Josephson current is absent for $\left|T_0 \right \rangle$, $\left|\uparrow \uparrow \right \rangle$, $\left|\downarrow \downarrow \right \rangle$ DQD ground states when $U \rightarrow \infty$~\cite{ArticleQDSplitting}.
In addition, we assume that $k_BT\ll J$, therefore we can neglect the thermal excitation of the triplet states. Here $J$ denotes the exchange interaction between the quantum dots (via superconducting electrodes) that can be also calculated with the 4-th order perturbation theory~\cite{ProbstPRB2016,ArticleQDSplitting}. Since we consider the superconducting electrodes with the superconducting gap $ \Delta $ in the quasiparticle density of states with assumption $ \Delta \gg k_B T_K $ one can neglect the effect of Kondo correlations on the value of the exchange interaction $J$.

\subsection{Empty quantum dots}

Let us first consider the regime in which the DQD is in the empty state~$\left|0\right\rangle$ ($\varepsilon_{\rm{u}},\varepsilon_{\rm{d}}>0$).
Josephson current consist of two components (local and nonlocal) $I_J=I_{\rm{local}}+I_{\rm{nonlocal}}$, which have a form:
\begin{widetext}
\begin{align}
I_{\rm{local}} &= {\left(I_{1\rm{u}}+I_{1\rm{d}}\right)}\sin \varphi \cos \phi_{\rm{AB}} \;,
\label{eqn:IJunspl0}\\
I_{\rm{nonlocal}}&= {I_2}\sin \varphi \cos \left(\phi_{\rm{Rlu}}+\phi_{\rm{Rru}}-\phi_{\rm{Rld}}-\phi_{\rm{Rrd}}\right)\;,
\label{eqn:IJspl0}
\end{align}
where:
\begin{align}
I_{1\delta} &= 4\frac{{2\rm{e}}}{\hbar }t_{\rm{l}}^2t_{\rm{r}}^2{N_{\rm{l}}}{N_{\rm{r}}}\int {d{\varepsilon _k}} \int {d{\varepsilon _q}} {u_k}{v_k}{u_q}{v_q}\frac{1}{{  {E_k} + {\varepsilon _{\delta}}}} \frac{1}{{ {E_q} + {\varepsilon _{\delta}}}}\frac{1}{{{E_k} + {E_q}}}\;,
\label{eqn:I10}\\
I_2 &= 4\frac{{2\rm{e}}}{\hbar }t_{\rm{l}}^2t_{\rm{r}}^2{N_{\rm{l}}}{N_{\rm{r}}}\int {d{\varepsilon _k}} \int d{\varepsilon _q}{u_k}{v_k}{u_q}{v_q}\left( \left( \frac{1}{E_k + \varepsilon _{\rm{u}}}\frac{1}{E_q + \varepsilon _{\rm{d}}} + \frac{1}{E_k + \varepsilon _{\rm{d}}} \frac{1}{E_q + \varepsilon _{\rm{u}}} \right)\frac{1}{E_k + E_q} \right.\nonumber\\
&\left. + \left( {\frac{1}{{{E_k} + {\varepsilon _{{\rm{u}}}}}} + \frac{1}{{{E_k} + {\varepsilon _{{\rm{d}}}}}}} \right)\left( {\frac{1}{{{E_q} + {\varepsilon _{{\rm{u}}}}}} + \frac{1}{{{E_q} + {\varepsilon _{{\rm{d}}}}}}} \right)\frac{1}{{{\varepsilon _{{\rm{u}}}} + {\varepsilon _{{\rm{d}}}}}} \right)\;,
\label{eqn:I20}
\end{align}
\end{widetext}
where $N_{\rm{l/r}}$ are the densities of states of the leads;
$u_k^2 = \left( {1 + \varepsilon _k/E_k} \right)/2$, $v_k^2 = \left( {1 - \varepsilon _k/E_k} \right)/2$, and ${E_k} = \sqrt {\varepsilon _k^2 + {\Delta ^2}}$.

We can regard $I_{\rm{local}}$ and $I_{\rm{nonlocal}}$
as currents related to the tunneling of unsplit and split Cooper pairs, respectively,
and, as it is apparent from Eqs.~(\ref{eqn:IJunspl0}) and~(\ref{eqn:IJspl0}),
only the local part of the Josephson current depends on $\phi_{\rm{AB}}$,
while the nonlocal component is affected by the Rashba phase.
If both electrons of an entangled pair, being in the singlet state $\left|S\right\rangle$,
travel through the same nanowire, their Rashba phases cancel due to opposite spins of the Cooper pair electrons,
and the Josephson current only depends on the AB phase.
If the flowing Cooper pair is split between both nanowires,
the AB phases of the electrons cancel, being opposite in the two nanowires;
as a result, nonlocal component of the Josephson current only depends on the Rashba phase, therefore we can observe the AC effect.

\subsection{DQD occupied by single electron}

Now we consider the regime in which the DQD is occupied by a single electron, for example QD$_{\rm{u}}$ is empty ($\varepsilon_{\rm{u}}>0$) and QD$_{\rm{d}}$ is occupied by a single electron ($\varepsilon_{\rm{d}}<0$). Calculated from Eq.~(\ref{eqn:IJE4}) Josephson current, consists of two components as in the previous case $I_J=I_{\rm{local}}+I_{\rm{nonlocal}}$, which have a form:

\begin{widetext}
\begin{align}
I_{\rm{local}} &= {\left(I_{1\rm{u}}-I'_{1\rm{d}}\right)}\sin \varphi \cos \phi_{\rm{AB}} \;,
\label{eqn:IJunspl01}\\
I_{\rm{nonlocal}}&= -{I'_2}\sin \varphi \cos \left(\phi_{\rm{Rlu}}+\phi_{\rm{Rru}}-\phi_{\rm{Rld}}-\phi_{\rm{Rrd}}\right)\;,
\label{eqn:IJspl01}
\end{align}
where:
\begin{align}
I'_{1\delta} &= 2\frac{{2\rm{e}}}{\hbar }t_{\rm{l}}^2t_{\rm{r}}^2{N_{\rm{l}}}{N_{\rm{r}}}\int {d{\varepsilon _k}} \int {d{\varepsilon _q}} {u_k}{v_k}{u_q}{v_q}\frac{1}{{  {E_k} - {\varepsilon _{\delta}}}} \frac{1}{{ {E_q} - {\varepsilon _{\delta}}}}\frac{1}{{{E_k} + {E_q}}}\;,
\label{eqn:I101}\\
I'_2 &=2\frac{{2\rm{e}}}{\hbar }t_{\rm{l}}^2t_{\rm{r}}^2{N_{\rm{l}}}{N_{\rm{r}}}\int {d{\varepsilon _k}} \int d{\varepsilon _q}{u_k}{v_k}{u_q}{v_q}\left( \left( \frac{1}{E_k + \varepsilon _{\rm{u}}}\frac{1}{E_k - \varepsilon _{\rm{d}}} + \frac{1}{E_q - \varepsilon _{\rm{d}}} \frac{1}{E_q + \varepsilon _{\rm{u}}} \right)\frac{1}{E_k + E_q} \right.\nonumber\\
&\left. + \left( {\frac{1}{{{E_k} + {\varepsilon _{{\rm{u}}}}}} + \frac{1}{{{E_q} - {\varepsilon _{{\rm{d}}}}}}} \right)\left( {\frac{1}{{{E_q} + {\varepsilon _{{\rm{u}}}}}} + \frac{1}{{{E_k} - {\varepsilon _{{\rm{d}}}}}}} \right)\frac{1}{{{E_k+E_q+ \varepsilon _{{\rm{u}}}} - {\varepsilon _{{\rm{d}}}}}} \right)\;,
\label{eqn:I201}
\end{align}
\end{widetext}

Comparing these expressions with those obtained for the empty DQD, one can notice that the local current now consists of two components with opposite signs (Eq.~(\ref{eqn:IJunspl0}) and (\ref{eqn:IJunspl01})) and the nonlocal current is negative. In the case of local current $I_{\rm{local}}$ the critical current is positive, when Cooper pairs flow through empty quantum dot and negative, when pairs go through occupied quantum dot. Therefore, change in the ground state of the quantum dot switches between 0 and $\pi$ junctions for local processes. This local current $I_{\rm{local}}$ is analogous to the Josephson current through a single quantum dot, in which the $\pi$-junction behavior has been observed~\cite{SpivakPRB91,vanDamNature06}. In general, the negative (positive) sign of the current contribution is a result of the odd (even) numbers of electron operators permutations~\cite{vanDamNature06}. The difference of factor 2 in equations (\ref{eqn:I10}) and (\ref{eqn:I101}) is due to the fact that for the empty quantum dot local Cooper pairs can flow in twice as many ways as it is in the case of the occupied quantum dot, where the flow of the Cooper pairs depend on the spin of the electron located on the quantum dot.

\subsection{DQD in the singlet state}

\begin{figure}[t!]
\centering
\includegraphics{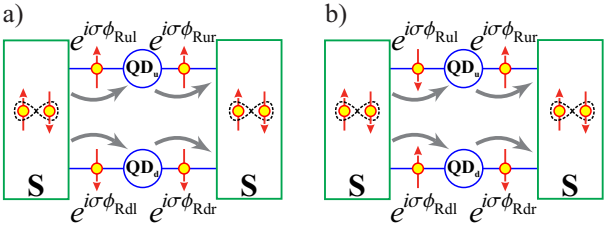}
\caption{Two types of the nonlocal processes: a) electrons tunneling from the quantum dots to the lead are followed and replaced by electrons with
the same spins, b) electrons tunneling to quantum dots have spins opposite to those of the preceding electrons.}
\label{fig:spinflip}
\end{figure}

An interesting effect can be observed in the singlet $\left|S\right\rangle$ ground state of the DQD ($\varepsilon_{\rm{u}},\varepsilon_{\rm{d}} <0$).
As in the previous cases, the Rashba spin-orbit interaction has no effect on the local part of the Josephson current.
The nonlocal current, however, involves two types of processes.
In the first type electrons tunneling from the quantum dots to the lead
are followed and replaced by electrons with the same spins (Figure~\ref{fig:spinflip}(a)).
In the other type of nonlocal current spin flip occurs:
electrons tunneling to quantum dots have spins opposite to those of the preceding electrons (Figure~\ref{fig:spinflip}(b)).
Spin flip occurs simultaneously at both quantum dots;
therefore, the tunneling process is elastic,
since the DQD ground state remains the same ($\left|S\right\rangle$).
Thus, in our system we have non-spin-flip and spin-flip nonlocal cotunneling processes,
\mbox{$I_{\rm{nonlocal}}=I_{\rm{nsf}}+I_{\rm{sf}}$},
and the Josephson current has the following components:
\begin{widetext}
\begin{align}
I_{\rm{local}} &= -{\left(I'_{1\rm{u}}+I'_{1\rm{d}}\right)}\sin \varphi \cos \phi_{\rm{AB}} \;,
\label{eqn:IJunsplS}\\
I_{\rm{nsf}}&= \frac{1}{2}{I''_2}\sin \varphi \cos \left(\phi_{\rm{Rlu}}+\phi_{\rm{Rru}}-\phi_{\rm{Rld}}-\phi_{\rm{Rrd}}\right)\;,
\label{eqn:IJsplSnsf}\\
I_{\rm{sf}}&= \frac{1}{2}{I''_2}\sin \varphi \cos \left(\phi_{\rm{Rlu}}-\phi_{\rm{Rru}}-\phi_{\rm{Rld}}+\phi_{\rm{Rrd}}\right)\;,
\label{eqn:IJsplSnsf}
\end{align}
where:
\begin{align}
I''_2 &=4 \frac{{2\rm{e}}}{\hbar }t_{\rm{l}}^2t_{\rm{r}}^2{N_{\rm{l}}}{N_{\rm{r}}}\int {d{\varepsilon _k}} \int d{\varepsilon _q}{u_k}{v_k}{u_q}{v_q}\left( \left( \frac{1}{E_k - \varepsilon _{\rm{u}}}\frac{1}{E_q - \varepsilon _{\rm{d}}} + \frac{1}{E_k - \varepsilon _{\rm{d}}} \frac{1}{E_q - \varepsilon _{\rm{u}}} \right)\frac{1}{E_k + E_q} \right.\nonumber\\
&\left. + \left( {\frac{1}{{{E_k} - {\varepsilon _{{\rm{u}}}}}} + \frac{1}{{{E_k} - {\varepsilon _{{\rm{d}}}}}}} \right)\left( {\frac{1}{{{E_q} - {\varepsilon _{{\rm{u}}}}}} + \frac{1}{{{E_q} - {\varepsilon _{{\rm{d}}}}}}} \right)\frac{1}{{{-\varepsilon _{{\rm{u}}}} - {\varepsilon _{{\rm{d}}}}}} \right)\;,
\label{eqn:I2S}
\end{align}
\end{widetext}
As a consequence, the Rashba phases can be adjusted
so that only $I_{\rm{nsf}}$ depends on the Rashba spin-orbit interaction,
$\phi_{\rm{Rlu}}-\phi_{\rm{Rld}}=\phi_{\rm{Rru}}-\phi_{\rm{Rrd}}$
(e.g. $\phi_{\rm{Rl} \delta}=\phi_{\rm{Rr} \delta}$, $\delta=\rm{u},\rm{d}$),
and $I_{\rm{sf}}$ is independent of both the AB and Rashba phases.

\section{Beating of the Josephson current}

\begin{figure}[t!]
\centering
\includegraphics{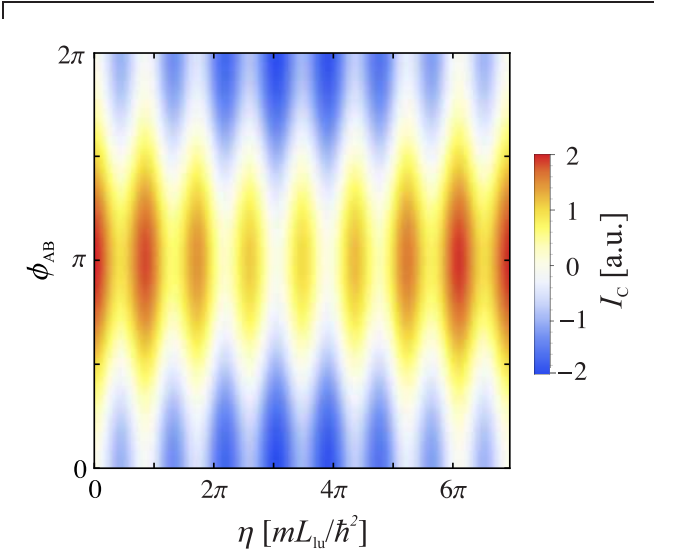}
\caption{Beating of the Josephson current: critical Josephson current~$I_{\rm{C}}$ versus Rashba parameter~$\eta$
for $\phi_{\rm{Rlu}}=m\eta L_{\rm{l u}}/\hbar^2$,
$\phi_{\rm{Rru}}=m\eta L_{\rm{r u}}/\hbar^2$, $L_{\rm{ru}}=1.3L_{\rm{lu}}$,
$\alpha=1/2$, $\phi_{\rm{Rld}}=\phi_{\rm{Rrd}}=0$, $\varphi=\pi/2$. }
\label{fig:spinFlip}
\end{figure}

Since the nonlocal component of the Josephson current for the singlet $\left|S\right\rangle$ ground state composes of two contributions:
the non-spin-flip $I_{\rm{nsf}}$ and spin-flip $I_{\rm{sf}}$, which differ in their Rashba phase dependence, we can observe beating of the Josephson current $I_{\rm{nonlocal}}$. Let us focus on the case with
$\phi_{\rm{Rld}} = \phi_{\rm{Rrd}}=0$ and $\phi_{\rm{Rlu}}\neq \phi_{\rm{Rru}}$.
We assume different lengths $L_{\rm{lu}}\neq L_{\rm{ru}}$ of the transport channel
on the left and right of the u-QD,
which implies different Rashba phases $\phi_{\rm{Rlu}/\rm{ru}}=m\eta L_{\rm{lu}/\rm{ru}}/\hbar^2$
($m$ denotes the electron mass), where the Rashba parameter $\eta$ can be tuned.
This results in a beat between currents $I_{\rm{nsf}}\propto \cos  \left(\phi_{\rm{Rlu}}+\phi_{\rm{Rru}}\right)$
and $I_{\rm{sf}}\propto \cos  \left(\phi_{\rm{Rlu}}-\phi_{\rm{Rru}}\right)$,
depicted in Figure~\ref{fig:spinFlip}.
Recently in InAs and InSb nanowires a large spin-orbit coupling was observed with effective spin-orbit length $l_{\rm{so}}\approx200$ nm and a Rashba parameter $\eta=0.2$ $\rm{eV}\cdot\rm{{\AA}}$~\cite{FasthPRL2007,NadjPergePRL2012,MourikScience2012,ManchonNatMat2015}. That indicate the length scale for $L_{\rm{lu}}$ and $L_{\rm{ru}}$, which makes the proposed effects possible to measure using present day technology.

\section{Cooper pair splitting efficiency}

We investigate how the presence of DQD affects the Cooper pair splitting efficiency. We focus on a symmetrical case, where $\varepsilon_{\rm{u}}=\varepsilon_{\rm{d}}=\varepsilon$ for $\left|0\right\rangle$ and $\left|S\right\rangle$ DQD ground state and $-\varepsilon_{\rm{u}}=\varepsilon_{\rm{d}}=\varepsilon$ for DQD occupied by a single electron. Equations~(\ref{eqn:I10}), (\ref{eqn:I20}), (\ref{eqn:I101}), (\ref{eqn:I201}), (\ref{eqn:I2S}) in this approximation have the form:

\begin{align}
I_{1\delta}&=2 I'_{1\delta} = 4\frac{{2\rm{e}}}{\hbar }t_{\rm{l}}^2t_{\rm{r}}^2{N_{\rm{l}}}{N_{\rm{r}}}\int {d{\varepsilon _k}} \int {d{\varepsilon _q}} {u_k}{v_k}{u_q}{v_q}\nonumber\\
&   \times \frac{1}{{  {E_q} + {\left|\varepsilon \right|}}} \frac{1}{{ {E_k} + {\left|\varepsilon \right|}}}\frac{1}{{{E_k} + {E_q}}}\;,
\label{eqn:I1simpl}\\
I_2&=I''_2 = 4\frac{{2\rm{e}}}{\hbar } t_{\rm{l}}^2t_{\rm{r}}^2{N_{\rm{l}}}{N_{\rm{r}}}\int {d{\varepsilon _k}} \int {d{\varepsilon _q}} {u_k}{v_k}{u_q}{v_q}\nonumber\\
&   \times \frac{1}{{ {E_q} + {\left|\varepsilon \right|}}}\frac{1}{{  {E_k} + {\left|\varepsilon \right|}}} \left( {\frac{1}{{{E_k} + {E_q}}} + \frac{1}{{{\left|\varepsilon \right|}}}} \right)\;,
\label{eqn:I2simpl}\\
I'_2 &= 2\frac{{2\rm{e}}}{\hbar }t_{\rm{l}}^2t_{\rm{r}}^2{N_{\rm{l}}}{N_{\rm{r}}}\int {d{\varepsilon _k}} \int d{\varepsilon _q}{u_k}{v_k}{u_q}{v_q} \nonumber\\
&  \times \left( \left( \frac{1}{E_k + \varepsilon }\frac{1}{E_k + \varepsilon } + \frac{1}{E_q + \varepsilon } \frac{1}{E_q + \varepsilon } \right)\frac{1}{E_k + E_q} \right.\nonumber\\
&\left. + \left( \frac{1}{E_k + \varepsilon } + \frac{1}{E_q + \varepsilon } \right)^2\frac{1}{E_k+E_q+2 \varepsilon } \right)\;.
\label{eqn:I201simpl}
\end{align}

We define Cooper pair splitting efficiency as:
\begin{equation}
\alpha = \frac{\left|I_{\rm{nonlocal}}\right|}{\left|I^{\rm{u}}_{\rm{local}}\right|+\left|I^{\rm{d}}_{\rm{local}}\right|+\left|I_{\rm{nonlocal}}\right|}\;,
\label{eqn:CPSeff}
\end{equation}
where $I^\delta _{\rm{local}}$ is the local Josephson current flowing through the $\delta$-quantum dot.

\begin{figure}[t!]
\centering
\includegraphics{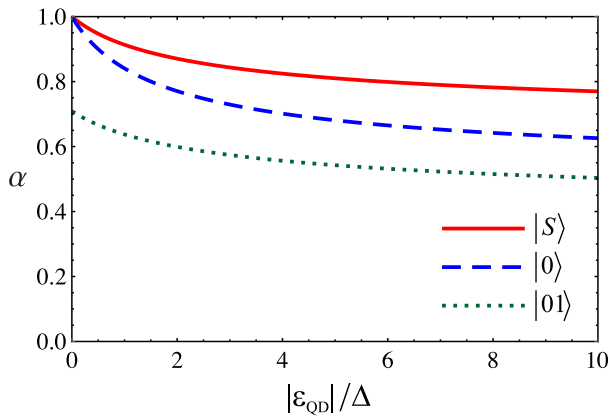}
\caption{Cooper pair splitting efficiency $\alpha$ in the absence of magnetic flux and Rashba spin-orbit interaction (no AB and AC effects) as a function of the quantum dot energy $\left|\varepsilon\right|$. Close to the resonance the splitting efficiency $\alpha\simeq 1$ for singlet and empty DQD ground state and $\alpha\simeq 0.71$ for DQD occupied by single electron. Out of resonance the efficiency is equal 2/3, 1/2, 2/5 for $\left| S \right \rangle$, $\left| 0 \right \rangle$, $\left| 0 1 \right \rangle$ ground state, respectively.}
\label{fig:efficiency}
\end{figure}

From Eqs.~(\ref{eqn:I1simpl}) and (\ref{eqn:I2simpl}), it follows that for singlet and empty dots ground state
close to resonance, $\left|\varepsilon_{\rm{\delta}}\right|/\Delta \ll 1$, the nonlocal Cooper pair current dominates,
since $I_2/I_1\propto \Delta/\left|\varepsilon_{\rm{\delta}}\right|\gg 1$. Close to resonance the efficiency is thus $\alpha\simeq 1$, while out of
resonance it tends towards 2/3 and 1/2 for $\left| S \right \rangle$ and $\left| 0 \right \rangle$ ground state, respectively.
For DQD occupied by single electron close to resonance $\alpha<1$ and tends towards some nonuniversal value, while out of resonance it tends towards 2/5.
The Cooper pair splitting efficiency as a function of the quantum dot energy $\varepsilon$ is shown in Figure~\ref{fig:efficiency}.

\section{Influence of Rashba phase on the DQD state}

In DQD Josephson junction Cooper pair can pass through the system using both channels in split or unsplit way, which leads to a different phase shift. If both electrons of Cooper pair (which is in spin singlet state) passes through the same (e.g up) quantum dot we have:
\begin{widetext}
\begin{equation}
\left|S\right\rangle \rightarrow \frac{1}{\sqrt{2}}\left(e^{i\left(\phi_{\rm{AB}}+\phi_{\rm{Ru}\mu}-\phi_{\rm{Ru}\mu}\right)}\left|\right\uparrow\downarrow\rangle-e^{i\left(\phi_{AB}+\phi_{\rm{Ru}\mu}-\phi_{\rm{Ru}\mu}\right)}\left|\right\downarrow\uparrow\rangle\right)=e^{i\phi_{\rm{AB}}}\left|S\right\rangle\;,
\label{eqn:singletUnsplittetRashba}
\end{equation}
and when electron pair passes in split way:
\begin{equation}
\left| S \right\rangle  \to \frac{1}{{\sqrt 2 }}\left( {{e^{i\left( {{\phi _{\rm{Ru}\mu}} - {\phi _{\rm{Rd}\mu}}} \right)}}\left| { \uparrow  \downarrow } \right\rangle  - {e^{i\left( {{\phi _{\rm{Rd}\mu}} - {\phi _{\rm{Ru}\mu}}} \right)}}\left| { \downarrow  \uparrow } \right\rangle } \right) =
\cos \left( {{\phi _{\rm{Ru}\mu}} - {\phi _{\rm{Rd}\mu}}} \right)\left| S \right\rangle  + i\sin \left( {{\phi _{\rm{Ru}\mu}} - {\phi _{\rm{Rd}\mu}}} \right)\left| {{T_0}} \right\rangle \;.
\end{equation}
\end{widetext}

Therefore due to the Rashba spin-orbit interaction, there is a possibility to manipulate the two electron state on the quantum dots inserted into two nanowires, both for single and two superconducting electrodes, and create admixture of triplet correlations $T_0$.

These entangled states on the double quantum dot can be detected, for example, by entanglement witnessing with ferromagnetic detectors~\cite{KlobusPRB2014,RozeAPPk2015} and by current measurements~\cite{BuszAPP2015,BuszPRB2017}.

\section{Critical current oscilations}
\begin{figure}[t!]
\centering
\includegraphics{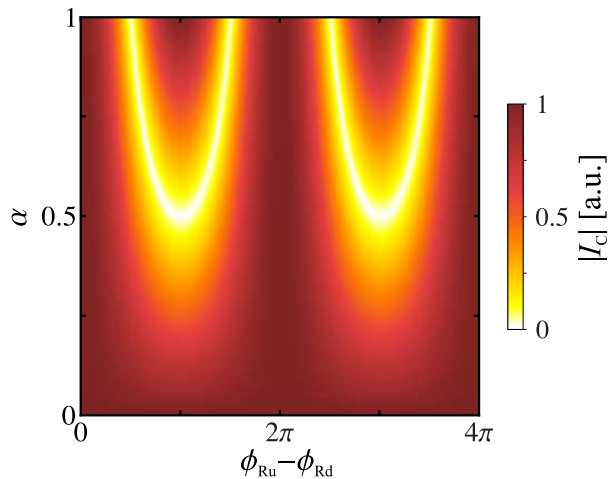}
\caption{Maximum Josephson current~$\left|I_{\rm{C}}\right|$ versus the Rashba phase $\phi_{\rm{Ru}}-\phi_{\rm{Rd}}$, as a function of splitting efficiency $\alpha$, $\phi_{\rm{AB}}=0$, $\phi_{\rm{Ru(d)}}=\phi_{\rm{Ru(d)l}}+\phi_{\rm{Ru(d)r}}$, $\varepsilon>0$. Change of the splitting efficiency $\alpha$ causes change of period of current AC oscillations.}
\label{fig:IcGamma}
\end{figure}

\begin{figure}[t!]
\centering
\includegraphics{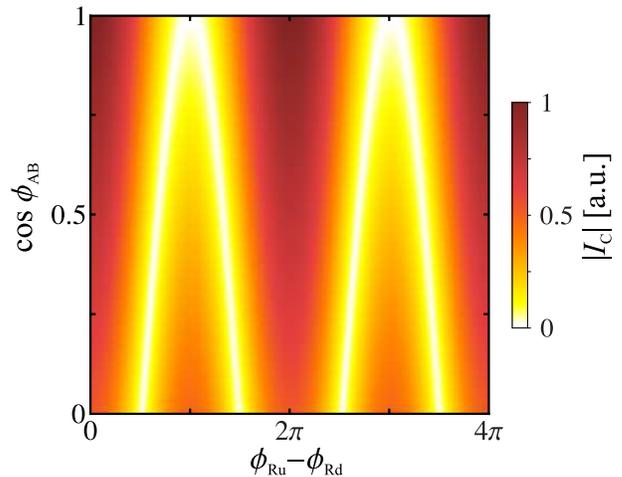}
\caption{Maximum Josephson current~$\left|I_{\rm{C}}\right|$ versus Rashba phase $\phi_{\rm{Ru}}-\phi_{\rm{Rd}}$, as a function of $\cos \phi_{\rm{AB}}$, for $\alpha=1/2$, $\phi_{\rm{Ru(d)}}=\phi_{\rm{Ru(d)l}}+\phi_{\rm{Ru(d)r}}$, $\varepsilon>0$. Change of the Aharonov-Bohm phase $\phi_{\rm{AB}}$ causes change of period of current AC oscillations.}
\label{fig:IcCosfAB}
\end{figure}

In this section we propose an additional method of
experimental confirmation of our predictions. It is based on
the measurement of the maximum Josephson current
$\left|I_{\rm{C}}\right|$, as commonly done experimentally.
Recent experiment by Deacon et al.~\cite{DeaconNC15} has shown that it
is possible to control experimentally the splitting of the Josephson
current and to distinguish between local and nonlocal component of the
Josephson current. The magnitude of the AC effect is directly related
to the splitting efficiency of the experimental setup, since Rashba
phase affects only the nonlocal Josephson current. Influence of the
splitting efficiency $\alpha$ on maximum Josephson current
$\left|I_{\rm{C}}\right|$ versus Rashba phase
$\phi_{\rm{Ru}}-\phi_{\rm{Rd}}$, $\phi_{\rm{AB}}=0$, for the junction
with the DQD in $\left | 0 \right \rangle$ ground state is shown in Figure~\ref{fig:IcGamma}. As can be seen from the plot, the AC effect is absent when only local component is present ($\alpha=0$). For $\alpha>0$ one can observe oscillations of the Josephson current related to the AC effect. In addition the period of the maximum current oscillations $\left|I_{\rm{C}}\right|$ doubles with increasing splitting efficiency $\alpha$. Change of the current oscillations period can be also controlled by tuning of the AB phase, as shown in Figure~\ref{fig:IcCosfAB}, for splitting efficiency $\alpha=\frac{1}{2}$. The same characteristics can be observed for the AB phase, as a function of $\alpha$ and $\cos\left(\phi_{\rm{Ru}}-\phi_{\rm{Rd}}\right)$. This also indicates presence of the local and nonlocal component. However now the AB oscillations are absent for $\alpha = 1$ and double the period when $\alpha = 0$.

\section{Conclusion}

We have studied Josephson junction with two parallel nanowires with a quantum dot in each wire. Due to the presence of the quantum dots one can increase the Cooper pair splitting efficiency. Our calculations confirm that the AC effect for Josephson supercurrent is possible for nonlocal split Cooper pairs in systems with unbroken time-reversal symmetry and it does not depend on the detailed geometry of the device. For local Cooper pairs the AC effect is absent, while the AB effect has the standard form.
For singlet ground state of the double quantum dot we have demonstrated that non-spin-flip and spin-flip transport processes
are related to different AC phases, which results in beating in the AC effect.
We have shown that the inserting of quantum dots in the two nanowires enables the manipulation of the two electron state on the quantum dots and create some triplet correlations $T_0$.

Recent experiments demonstrated a large spin-orbit coupling in InAs and InSb nanowires with effective spin-orbit length $l_{\rm{so}}\approx200\ \rm{nm}$ and a Rashba parameter $\eta=0.2\ \rm{eV}\cdot\rm{{\AA}}$~\cite{FasthPRL2007,NadjPergePRL2012,MourikScience2012,ManchonNatMat2015}. It was also shown that there is a possibility of: assembling two Rashba parallel InAs nanowires with quantum dots (the length $\approx 250\ \rm{nm}$ and the
distance between nanowires $\approx 100\ \rm{nm}$)~\cite{BabaAPL2017}; fabrication of the Josephson junction with $\approx 200\ \rm{nm}$
long InSb Rashba nanowire with quantum dot, with spin-orbit length
$l_{\rm{so}}\approx350\ \rm{nm}$~\cite{SzombatiNaturePhys2016}; fabrication InSb nanowire "hashtags" (rectangular loops) that can be connected to
superconducting electrodes~\cite{GazibegovicNature2017}. These
examples of experimental work suggest that the proposed effects can be
detected using the present day technology. In
particular, it should be possible to observe oscillations of the maximum Josephson current as a function of the Rashba phase.

\section{Acknowledgments}

We would like to thank J.~Barna\'{s}, M.~Braun, B.~Braunecker, F.~Dominguez,
T.~Kontos, J.~K\"onig, T.~Martin, C.~Sch\"{o}nenberger, B.~Sothmann, J.~Tworzyd{\l}o, and A.~L.~Yeyati for helpful discussions.
D. T., P. B. and J. M. received support from the EU FP7 Project SE2ND (No. 271554) and the National Science Centre, Poland, grant 2015/17/B/ST3/02799. R. L. was supported by MINECO Grants FIS2014-52564 and MAT2017-82639. R.~\v{Z}. acknowledges the support of the Slovenian Research Agency (ARRS) under Program P1-0044 and J1-7259. M. L. was supported by the National Research Foundation of Korea (Grant Nos. 2011-0030046).

\end{document}